\documentclass[showpacs,twocolumn,prl,linenumbers]{revtex4}%

\usepackage{amsmath,graphicx}
\usepackage[dvipsnames]{xcolor}\usepackage[colorlinks=true,citecolor=blue]{hyperref}
\usepackage{ulem}

\begin{document}

\title{Long- to short-junction crossover and field-reentrant critical current in Al/Ag-nanowires/Al Josephson junctions}

%{Magnetic-field-dependent reentrance of the critical current of Al/Ag-nanowires/Al Josephson junctions}
%\Al/Ag-nanowires/Al Josephson junctions:  Magnetic field-dependent reentrance of the critical current, and full range from short to long junction}
%Al/Ag-nanowires/Al Josephson junctions:  Magnetic field-dependent reentrance of the critical current, and full length of the supercoonducting proximity effect.

\date{\today}
\author{A. Murani$^{1,2}$, S. Sengupta$^{1,3}$,   A. Kasumov$^{1,4}$, R. Deblock$^{1}$, Caroline Celle $^{5}$, J-P. Simonato $^{5}$, H. Bouchiat$^{1}$, and S. Gu\'eron$^{1}$ }

\affiliation{$^{1}$ Laboratoire de Physique des Solides, Univ. Paris-Sud, CNRS, UMR 8502, Universit\'e Paris-Saclay, 91405 Orsay Cedex, France}
\affiliation{$^{2}$ Quantronics Group, Service de Physique de l’Etat Condens\'e (CNRS UMR 3680), IRAMIS, CEA-Saclay, 91191 Gif-sur-Yvette, France. }
\affiliation{CSNSM, Univ. Paris-Sud, CNRS/IN2P3, Universit\'e Paris-Saclay, 91405 Orsay Cedex, France}
\affiliation{$^{4}$ IMT RAS, 142432 Chernogolovka, Moscow region, Russia. }
\affiliation{$^{5}$ Univ. Grenoble Alpes, CEA, LITEN DTNM, F-38054 Grenoble, France. }
\begin{abstract}

We have probed the superconducting proximity effect through long high-quality monocrystalline Ag nanowires, by realizing Josephson junctions of different lengths, with different superconducting materials. Thanks to the high number of junctions probed, both the contact resistance and electron diffusion constant could be determined, enabling a comparison of the measured critical current  to theoretical expectation, over the entire regime from short to long diffusive junction. Although the length dependence of the critical current is as expected, the amplitude of the $R_{N}I_c$ product is smaller than predicted by theory. We also address the magnetic field dependence of the critical current. The quasi-gaussian decay of the critical current with field expected of a long narrow junction is observed for all superconducting contacts we used except for aluminum. We present the striking non-monotonous effect of field on the critical current of junctions with aluminum contacts, and analyze it in terms of improved quasiparticle thermalization by a magnetic field.

\end{abstract}
\maketitle

\section{Introduction}

The superconducting proximity effect, in which a superconductor (S) confers superconducting-like properties to a non-superconducting (also called "normal", N), quantum coherent, material connected to it, is epitomized by the fact that a supercurrent can flow through an S/N/S junction. The critical  current, or maximal supercurrent, is a measure of the characteristic energy of the junction. It is related either to the superconductor's energy gap $\Delta$, or to the Thouless energy, associated to the inverse dwell time in the normal metal: $R_{N}I_{c} \propto \rm min({\Delta, \hbar D_N/L^2}) $ \cite{Dubos}, where $\rm D_N$ is the diffusion constant and L the length of the normal segment. This rule for the figure of merit of a proximity junction has only been tested so far for diffusive S/graphene/S junctions \cite{grapheneChuan}. In that work, the tunability of graphene was used to vary $\rm D_N$ with a gate voltage in order to probe the relation. In this paper, rather than varying the diffusion constant, we rely on high quality monocrystalline Ag wires of remarkably constant properties \cite{Simonato,Cheng,Mayousse} (diameters, diffusion constant, contact resistance) to probe this relation. To this end we have measured the supercurrent induced through almost twenty Ag nanowire segments, whose  lengths range between 200 nm and 5 microns, sampling the regimes from short to long SNS junctions, with an almost three order of magnitude range of the Thouless energy. In contrast to the S/graphene/S junctions, that were several micrometers wide, the S/Ag/S junctions are narrow, with a Ag wire diameter of roughly 50 nm.  This has several consequences, in particuar the field dependence of the critical current is drastically different. For most superconducting contacts we tried (Pd/Nb,W,Pd/ReW), we find a monotonously decaying critical current as the magnetic field is increased, as is expected of long narrow junctions \cite {Montambaux,Chiodi,Cuevas}. For Al contacts however we find that the magnetic field increases the critical current at low fields. We interpret this in terms of  a positive role of magnetic field for quasiparticle thermalization, a sensitive issue in narrow metallic wires. 

\section{Sample Fabrication}
\begin{figure}[h!]
	\hspace*{-1.0cm}
	\centering
	\includegraphics[width=10 cm] {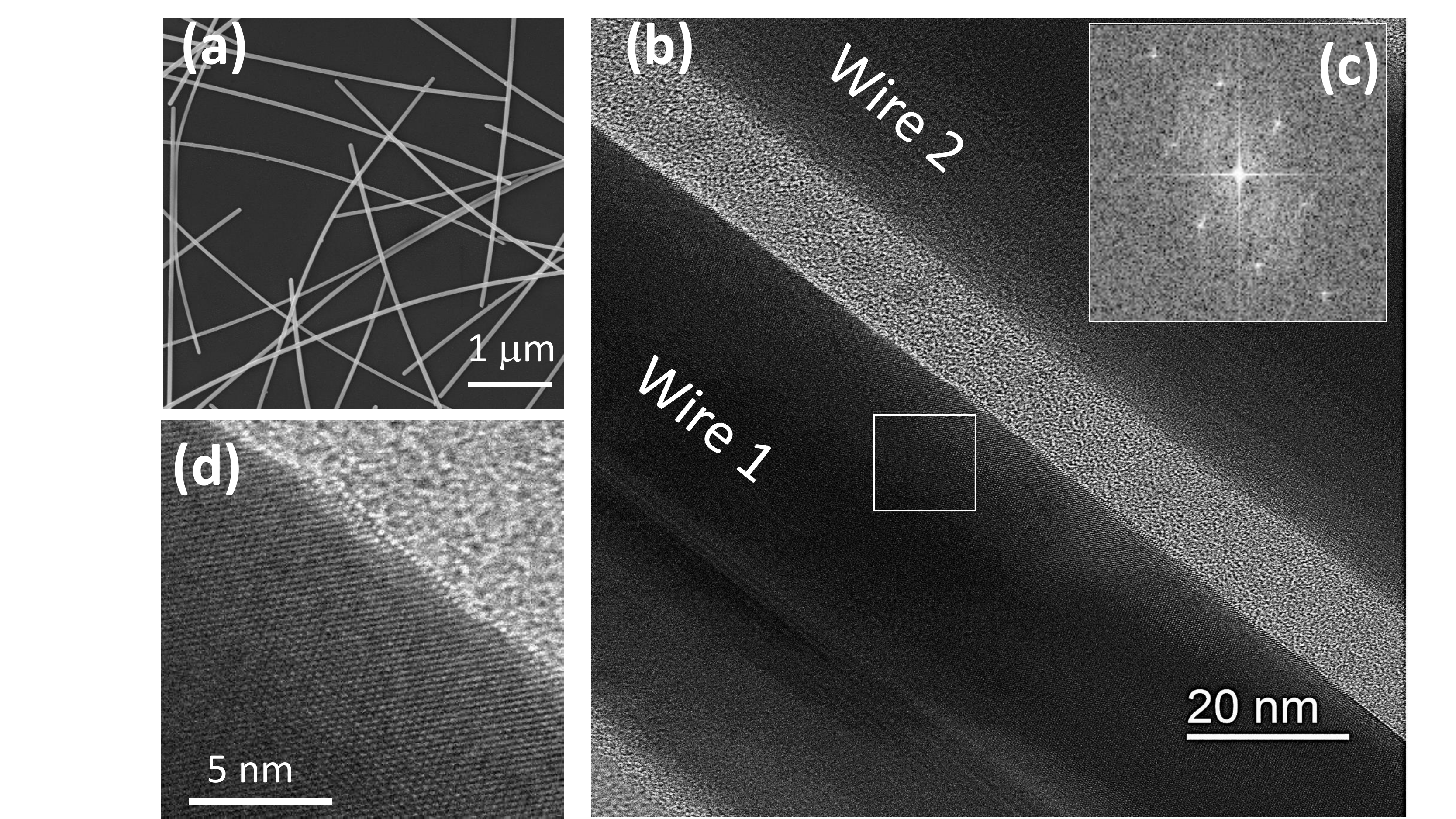}
	\caption{Imaging the Ag nanowires. (a) Scanning electron micrograph (SEM). (b) Bright field Scanning Transmission electron micrograph (STEM) of two Ag nanowires such as those measured. The Fast Fourier transform of the zone indicated by the white square is shown in (c), demonstrating the high crystallinity of the wires. Interestingly, the two parallel wires, labeled Wire 1 and Wire 2, have the same crystalline orientation (not shown). (d) Higher-magnification, atomic-resolution, STEM image.}
	\label{TEM}
\end{figure}

The synthesis of Ag nanowires was performed according to a recently published protocol, developed for the production of transparent electrodes based on nanowire random networks  \cite{Simonato}. In order to enhance induced superconductivity through the wires, a highly purified source of $\rm AgNO_3$ (purity $99.9999 \%$) was used in the present synthesis. Briefly, the synthesis consists in the reduction of silver salt through a polyol process and the purification is realized by a two-step decantation procedure. The sheet resistance of a spraycoated network of these nanowires was $20~\Omega/ \rm sq $ at $90 \%$ transparency (measured at 550 nm). The nanowire diameter was typically 50 nm, as determined by STEM and SEM characterizations. STEM observation are also compatible with a five-fold symmetry and a highly strained core.

 To isolate individual nanowires, a drop of a highly diluted solution of these nanowires in VLSI methanol is deposited on an oxide covered silicon substrate. Nanowires are selected in an optical microscope, using a polarizer to enhance contrast.
Contacts are defined by electron-beam lithography, and the superconducting materials are deposited by either e-gun evaporation (Ti/Al with thickness 5 nm/100 nm), dc sputtering (Pd/Nb or Pd/ReW , thicknesses 6 nm/100 nm), or focused-ion-beam (FIB)-assisted deposition (W-based compound, of thickness about 100 nm). The quality of the contacts between the silver nanowire and the superconductor is enhanced by a prior in-situ Argon ion etching step in the case of e-gun or sputter deposition. FIB-assisted deposition, by contrast, requires no additional etching step, for W deposition is induced by Ga-ion-assisted decomposition of a hexacarbonyl W gaz and is concommittant to a slight etching of the Ag wire by the Ga ions.  In fact, we find that FIB-assisted-deposition of superconducting contacts is particularly suited to contact wires covered by an insulating oxide \cite{BismuthChuan}. In the following we focus on Ag nanowires connected to Ti/Al electrodes, since they produce the most reliable results and the striking field dependence that we wish to address.

 \section{Mean free path and contact resistance determination}
\begin{figure}[h!]
	\centering
	\includegraphics[width=\linewidth]{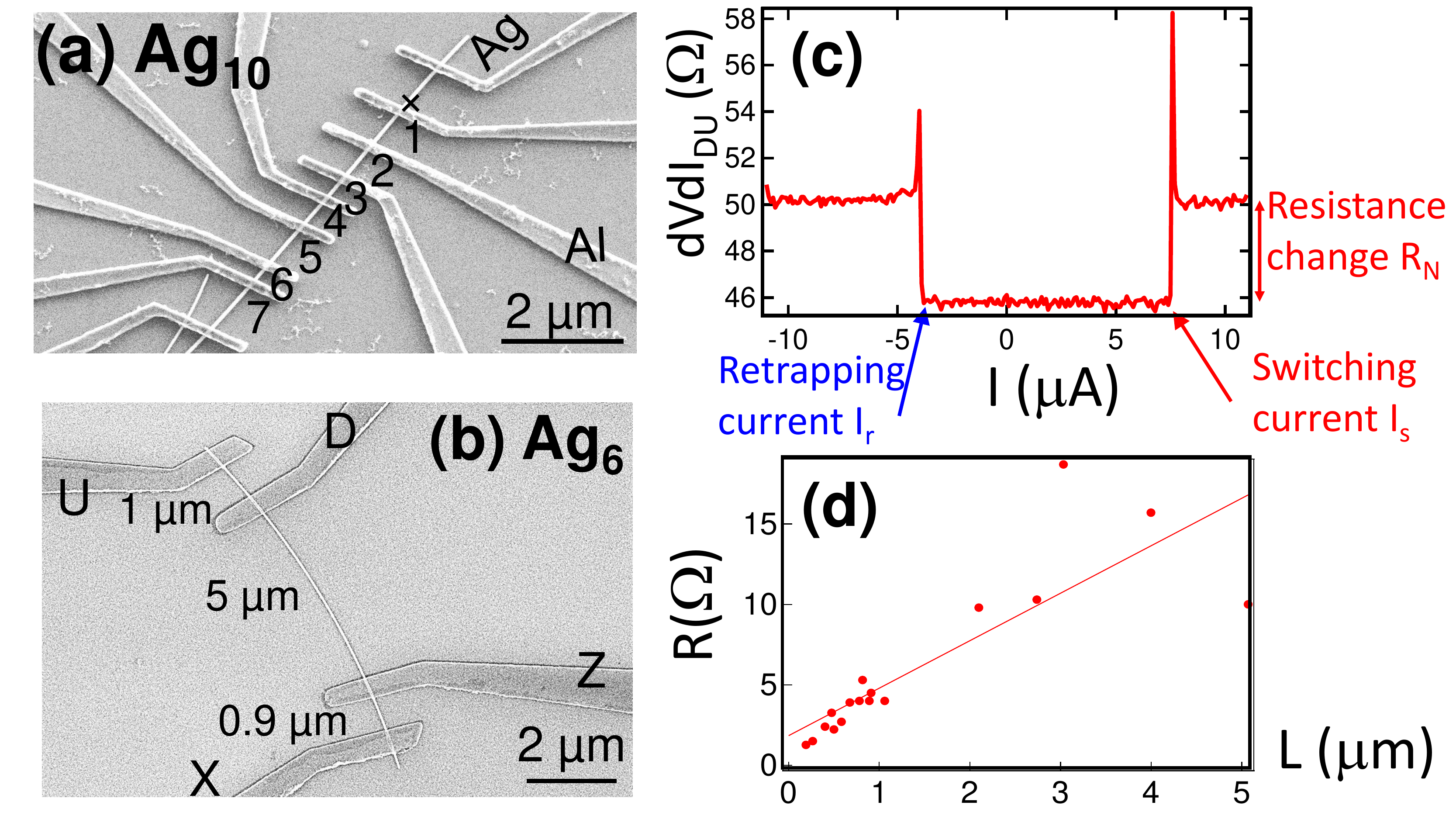}
	\caption{Diffusive transport in Ag nanowires. The Scanning Electron Micrograph of two of the wires, Ag6 and Ag10, with TiAl contacts, are displayed in (a) and (b). (c) Differential resistance as a function of dc current of sample TiAl Ag6-DU, measured at $\rm T\simeq 100~mK$, in a two wire configuration. The $46~\Omega$ resistance corresponds to the series resistance of the wires running from  room temperature down to the sample at mK temperature. As the current is swept from negative to positive, the resistance of the segment itself is zero for a current between $-4 ~\rm \mu A$, the retrapping current $\rm I_r$, and $7.5 ~\rm \mu A$, the switching current $\rm I_s$. The resistance change of $4~\Omega$ at the switching current is the normal state resistance of the segment $\rm R_N$. (d) Normal resistance of 17 segments from eight Ag nanowires with Ti/Al superconducting contacts. The linear increase of resistance with segment length indicates a diffusive behavior from 100 nm to 5 microns. The contact resistance and resistance per micron we deduce from a 	 linear fit to the data (continuous line) is $1.8 \pm 1 ~\Omega$, yielding a resistivity of $6~\pm 2~10^{-9} ~\rm \Omega.m $, assuming that all the wires have a diameter of 50 nm. This resistivity is similar to that measured in \cite {Cheng}.}
	\label{Ohm}
\end{figure}

The wire resistivity and contact resistances were determined by probing several Ag segments of different lengths with the superconducting proximity effect. Indeed the resistance jump as the junctions switches from a (proximity-induced) superconducting state with zero resistance to a resistive state as the current exceeds the critical current (see Fig. \ref{Ohm}), gives the normal resistance $R_N$ of the segment.
Plotting $R_N$ as a function of the Ag segment length thus yields the parameters characterizing the transport regime (contact resistance and elastic mean free path). Figure \ref{Ohm} displays the variation of the normal state resistance with segment length of 17 segments from eight Ag nanowires, connected to Ti/Al superconducting contacts. The linear dependence indicates that the transport regime is diffusive, and yields the contact resistance, mean free path, and wire resistivity. Indeed the length dependence of the resistance can be written as $R(L)=2R_c+\frac{R_{Q}}{M}\frac{L}{l_e}$, where $R_c$ is the contact resistance at each interface, $R_Q=h/2e^2$ the resistance quantum for a spin degenerate channel, L the wire length, $l_e$ the elastic mean free path, and $M$ the number of channels. Twice the contact resistance is the sum of the Sharvin resistance \cite{Sharvin} contribution of a partially transmitted channel of transmission $\tau$: $2R_c=\frac{R_Q}{M} (1+\frac{1-\tau}{\tau})=\frac{R_Q}{M}\frac{1}{\tau}$, so that $R(L)=\frac{R_Q}{M}(\frac{1}{\tau}+\frac{L}{l_e})$ \cite{Datta}.
 In a 3D system of section S and Fermi wavevector $\rm k_F$, $M=k_F^2S/(4\pi)$. Given the diameter d=50 nm and Ag's Fermi wavevector $k_{\rm F}=1.2~ 10^{10}~\rm m^{-1}$, we find $M \simeq 2 ~10^4$ channels. 
The slope of $3~ \Omega/\mu \rm m$ thus corresponds to a resistivity $\rho=(R-2R_c)S/L\simeq 6~10^{-9}~\rm \Omega.m$ and a mean free path $l_e=\frac{R_Q/M}{(R-2R_c)/L}\simeq \rm 200~nm$. 
The extrapolation to zero length yields a contact resistance of $2~ \rm \Omega$ for TiAl contacts (and thus $\tau \simeq 0.2$),  but this contact resistance varies from 0 to $10~\rm \Omega$ depending on the contact material.
The origin of disorder is not known, but surface scattering as well as the highly stressed core of the nanowire are likely to play a role. Twinning boundaries, by contrast, are not expected to cause much elastic scattering \cite{Bid}.

\section{Full range of proximity effect, from short to long junction}
The superconducting proximity effect occurs at low enough temperature that the contacts are in the superconducting state, and that the Ag nanowire is quantum coherent over the entire segment length between two superconducting contacts. Andreev bound states, coherent superpositions of electron and hole-like quasiparticles, then form in the normal metal, and can shuttle pairs from one superconductor to the other through the normal wire. This leads to a dissipationless supercurrent  through the junction, whose maximum value is the critical current $I_c$.
\begin{figure}[h!]
	\centering
	\includegraphics[width=\linewidth]{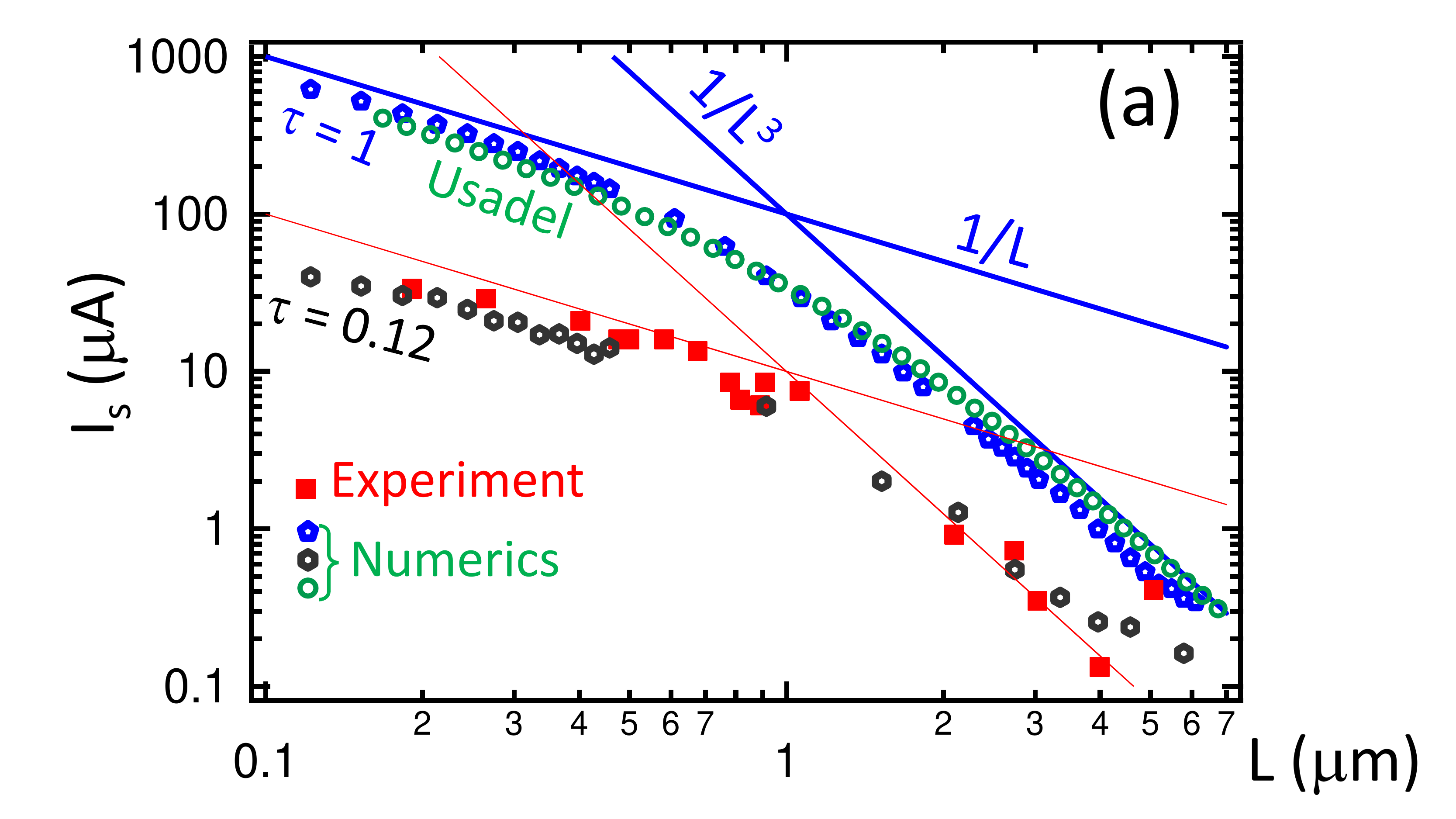}
	\includegraphics[width=\linewidth]{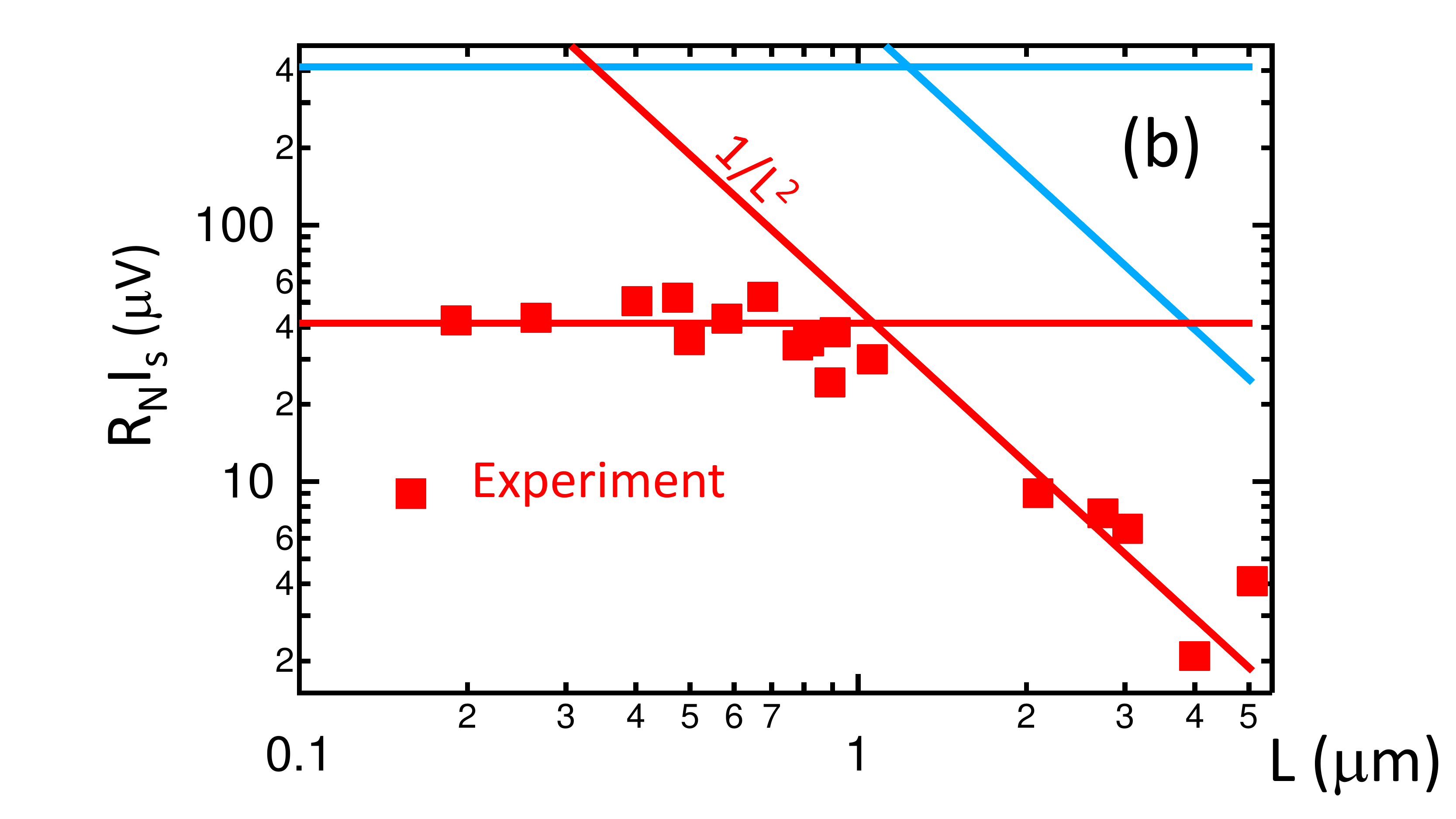}
	\caption{Full range of the proximity effect, from short to long junction regime, in Ag nanowires with TiAl contacts. (a) Switching current as a function of length for the 17 segments (full red squares). Open black and blue symbols are the numerically calculated length dependence of the critical current of a diffusive SNS junction for interface transmissions $\tau=0.12$ and $\tau=1$ respectively. Those results are  obtained  from the diagonalisation of the Bogolubov-de Gennes Andreev spectrum of a tight binding Hamiltonian on a square lattice with on-site disorder. Open green symbols are the results of the Usadel theory with perfect transmission, from \cite {Dubos}. In the simulations, the amplitude of the disordered potential and the superconducting gap	where chosen such that $l_e\simeq \xi_s/2=6a$ and the sample  length was varied between $L=0.25 \xi_s$ and $L=30~\xi_s$, which roughly corresponds to the experimental range. Interestingly, in this parameter range the Josephson current amplitude varies linearly with $\tau$ down to $\tau \simeq ~10^{-2}$. The transmission $\tau=0.12$ which can reproduce the experimental data is notably different from our estimate of $\tau \in [0.23,0.8]$ for the contact resistance to the Ag wire (see text).  (b) $\rm R_NI_s$ as a function of length. Plot (b) shows how the $\rm R_NI_s$ product is constant in the short junction regime, and decays as $1/L^2$ in the long junction regime, as it should. (b) also compares the experimental values to theory for perfect transmission (blue lines), and demonstrates how the experimental values are systematically smaller. The crossover between short and long junction regimes, however, occurs at the predicted length.  }
	\label{Proximity}
\end{figure}
In this section we characterize the proximity effect in each junction by the switching current, i.e. the current above which the S/Ag/S junction switches from non resistive to resistive. This switching current is expected to be not much smaller than $I_c$ when temperature is smaller than 
the Josephson energy $E_J=\frac{\hbar}{2e}I_c$. This is the regime in the samples discussed in this paper, for which the switching currents are high (above $\rm 100~nA$).
 We current bias the junction and use a standard lock-in technique to measure the differential resistance, see sketch in Fig.  \ref{Ohm}.  The switching current is plotted as a function of segment length in Fig.  \ref{Proximity}a. Two regimes are clearly seen, that correspond to two different powerlaw dependences of the switching current with length.  The first is the regime of short junctions, in which the junction is shorter than the superconducting coherence length $\xi_{\rm S}^{\rm N} = \sqrt{\hbar D_{\rm N} /\Delta}$, with $D_{\rm N}$ the diffusion constant in the Ag wire and $\Delta$ the superconducting gap. Given the gap of Al $\Delta \simeq 0.2 ~ \rm  meV$ and the diffusion parameters in Ag estimated earlier, $D_{\rm N}=v_{\rm F}l_{\rm e}/3 = 0.032 \rm m^2/s$, this yields $\xi_{\rm S}^{\rm N} \simeq 0.36~ {\rm \mu m}$.   In that regime, the superconducting gap gives the energy scale of most phenomena.
In the opposite regime of long junction, in which the junction is much longer than $\xi_{\rm s}$, it is the Thouless energy $E_{\rm Th} = \hbar D_{\rm N}/L^2$ that governs the physics.
Given the relation $R_{N}I_{c} \propto \rm min({\Delta, \hbar D_N/L^2}) $, in the diffusive regime the critical current scales as $I_{\rm c}\sim 1/L$ in the short junction limit, and $I_{\rm c}\sim 1/L^3$ for the long junction limit. These two different power laws are clearly displayed in Fig. \ref{Proximity}a. The crossover from short junction to long junction is located at $L_{\rm cr}=1\,{\rm \mu m}$, and is theoretically given by $\sqrt{10.82\hbar D_{\rm N}/2.07\Delta}$, which is roughly twice the coherence length in the normal metal $\xi_{\rm S}^{\rm N}=\sqrt{\hbar D_{\rm N}/\Delta}$. And indeed, the coherence length we extract from the crossover, $\xi_{\rm S}^{\rm N,exp} = 0.43~\rm \mu m$, is close to the $\xi_{\rm S}^{\rm N}\simeq 0.36~ {\rm \mu m}$ estimated above. 
Rather than plotting the switching current as a function of length, Fig.  \ref{Proximity}b plots the so-called junction "figure of merit", or $R_{N}I_s$ product, which, according to the theory of the proximity effect in the case of perfectly transparent contacts should be constant and equal to $2.07~\Delta$ in the short junction regime, and to $10.82~E_{\rm Th}\simeq 1/L^2$ in the long junction regime \cite{Kulik,Likharev,Dubos}. 
 The switching current (defined in Fig. 2c) shown in Fig.  \ref{Proximity}b qualitatively follows the theoretical prediction for $I_c$, both in terms of the length dependences and the crossover length between short and long junction (around one micrometer). Quantitatively however, the measured $R_{N}I_s$ is roughly five times smaller than the predicted $R_{N}I_c$ in both limits.

This reduction is probably due to the imperfect transmission at the N/S interface, as explained in \cite{Cuevasinterface}. However, the reduction we find here, just like the reduction found in S/graphene/S junction \cite{grapheneChuan}, does not agree quantitatively with the prediction of the Usadel theory used in \cite{Cuevasinterface}. In the S/graphene/S case, the transmission deduced from the contact resistance was estimated to be $\tau= 0.15\pm 0.05$. In the present case of Ag connected to TiAl contacts, the contact resistance of $2R_c=1.8 \pm 1 ~\Omega$ determined in the previous section, translates, given the roughly $2~10^4$ conduction channels in the nanowires, into $\tau \in [0.23,0.8]$. As in \cite{grapheneChuan}, we have conducted tight binding simulations \cite{Simus} of a few-channel diffusive wire connected to superconducting contacts, with varying interface transmission, for different wire lengths. We find that the reduction factor in $R_NI_c$, as well as the entire length dependence are well reproduced by the numerical calculations for a transmision of 0.12, as shown in Fig. 3 a.
This transmission is significantly smaller than the smallest value compatibe with the measured contact resistance.  Therefore, in addition to a bad contact transmission (tyical of contacts between metals and semiconductors or carbon-based materials such as graphene), other physical mechanisms are needed to explain the reduction of $R_{\rm N}I_{\rm c}$. 
One possibility is a reduced induced superconductivity due to the small size of the superconducting Al contacts with respect to the superconducting coherence length of Al films: $\xi_{\rm Al} =\sqrt{\xi_{0,\rm Al} l_{\rm e, Al}}\simeq 600~{\rm nm}$ where $\xi_{0, \rm Al}=\hbar v_{\rm F}/\Delta_{\rm Al} \simeq 5\,{\mu m}$ is the coherence length of clean Al, and $l_{\rm e,Al}$ is estimated using a thickness-limited mean free path in our Al films $l_{\rm e,Al}=t=75\,{\rm nm}$. 
Another possibility is the effect of out-of-equilibrium quasiparticles in the aluminum contacts. Their strong influcence on the field-dependence of the switching current will be discussed in the next section.
We note that such a discrepancy between calculated and measured $R_NI_c$ was not observed in Al/Au/Al junctions measured by Angers et al. \cite{Angers}, in which the Al wires had a larger section and the S/N contact area was much larger.

\section{Field dependence of the Proximity effect: standard gaussian decay in most junctions, and surprising reentrance features in Al/Ag/Al junctions}
\begin{figure}[h!]
	\centering
	\includegraphics[width=\linewidth]{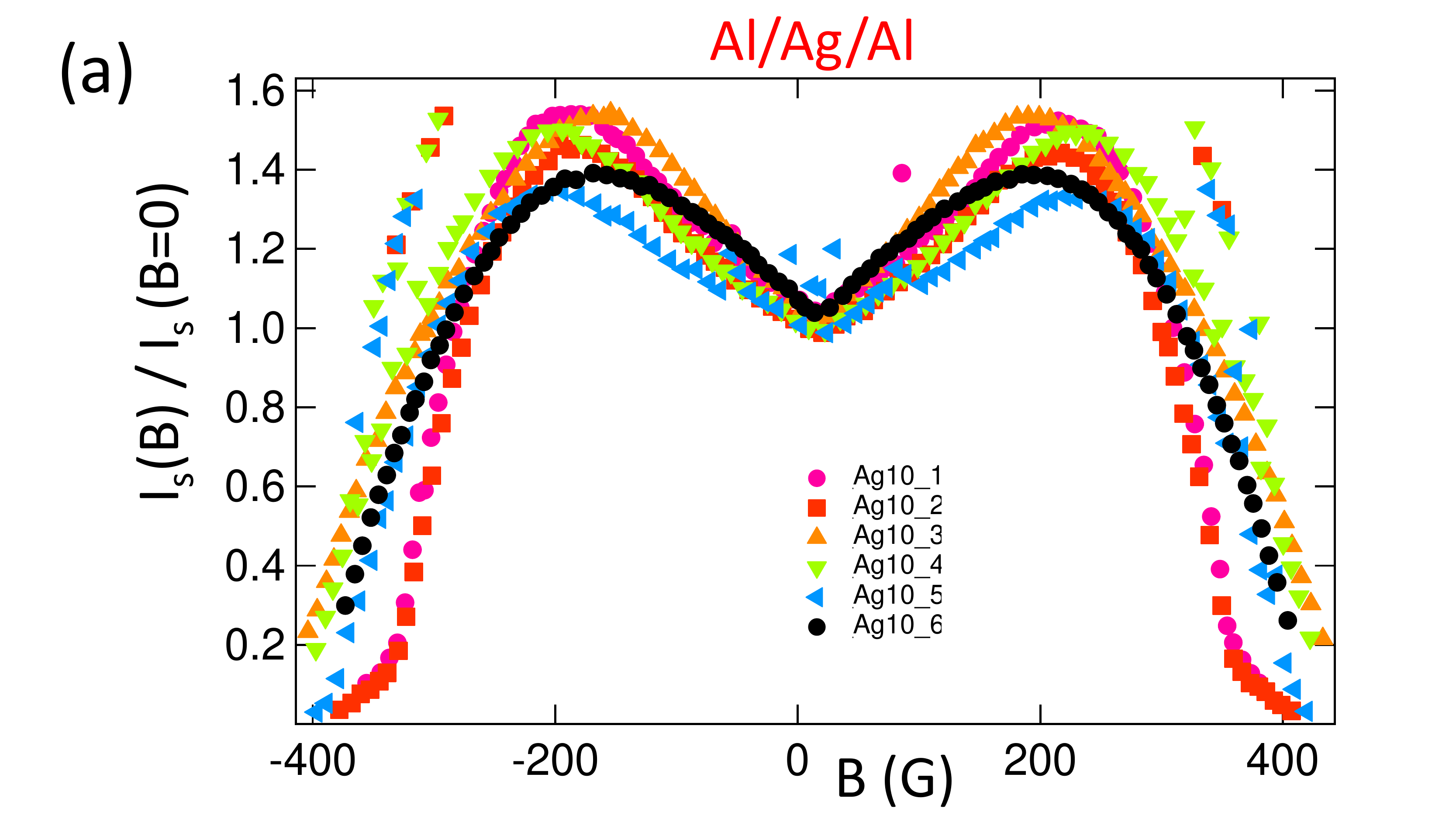}
		\includegraphics[width=\linewidth]{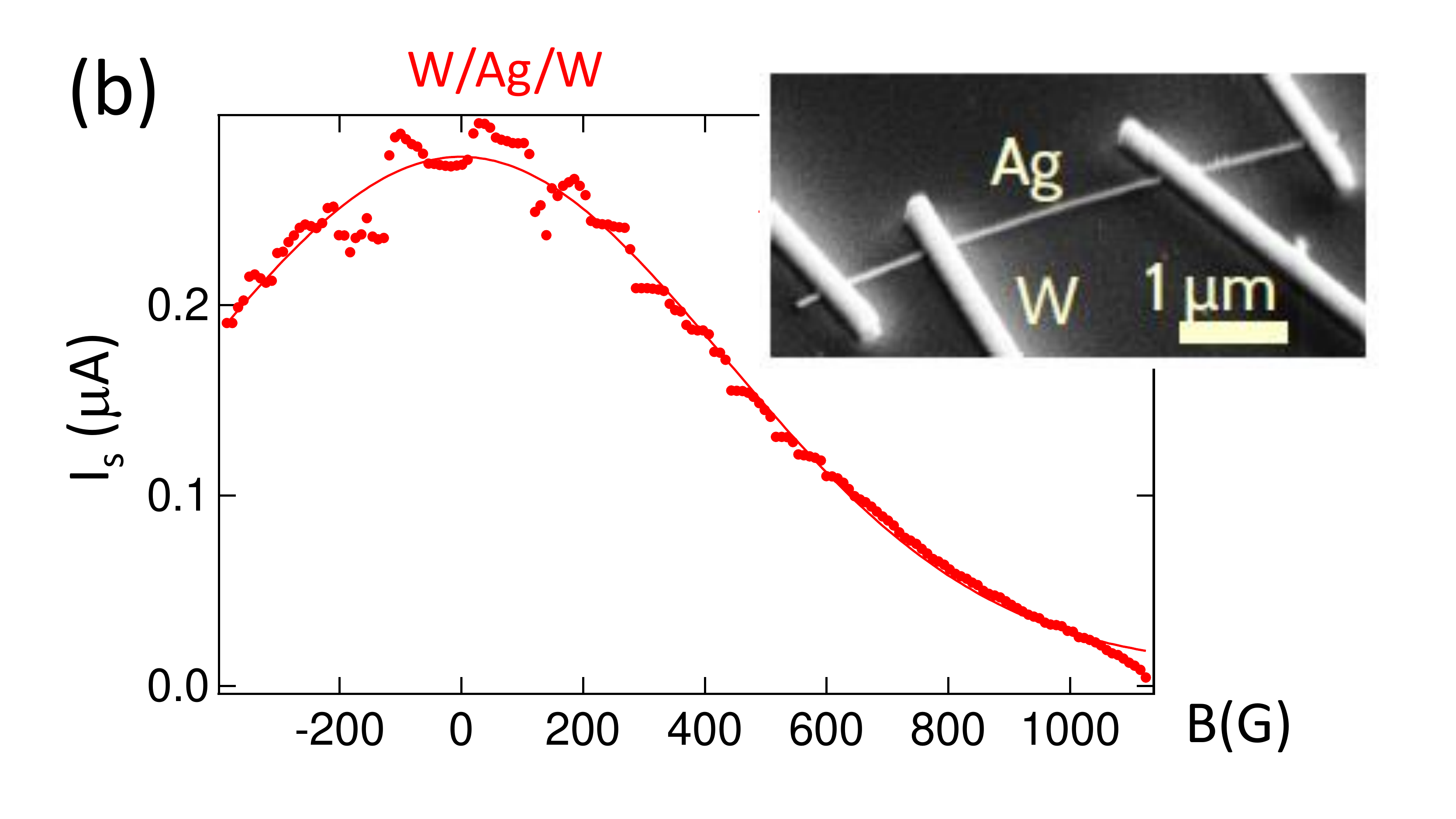}
	\caption{Comparison of the effect of magnetic field on the critical current for W/Ag/W and Al/Ag/Al junctions, measured in a four wire configuration. Whereas the switching current of the Al/Ag/Al junction (A) first increases with field before decaying, the W/Ag/W junction (in B, SEM image inset) display a monotonous, gaussian-like decay with magnetic field, as theoretically expected for unidimensionnal geometry. The continuous line is a gaussian fit.}
	\label{Reentrance_comp}
\end{figure}

The effect of a magnetic field on an SNS junction depends on the junction geometry, specifically on the aspect ratio of the N part. Whereas wide two-dimensional-like junctions exhibit a non monotonous, Fraunhoffer-like critical current versus field pattern, narrow one-dimensional-like junctions display a monotonous, gaussian-like decay with field of the critical current \cite {Montambaux, Chiodi, Cuevas}. Such a gaussian-like decay is demonstrated in Fig. \ref{Reentrance_comp}b, for the Ag nanowire between FIB-grown W electrodes. It was also reported in 1D Nb/Au/Nb SNS junctions where the Au nanowires were made of polycristalline thin films \cite{Chiodi}.

In striking contrast, the field dependence of the critical current through S/Ag/S junctions with Al contacts, measured in four wire (4W) configuration, displayed in Fig. \ref{Reentrance_comp}a, is non monotonous. It first increases with field, then decreases in a gaussian manner above about 150 G. One of the main results of this paper is that all junctions with Al contacts display such a behavior, as also shown in the appendix.  This reentrant behaviour was also reported previously for Al/Au/Al junctions (but not Nb/Au/Nb junctions), with a relative $10 \%$ effect, in which the Au wires were made of polycristalline thin films, roughly 100 nm wide and 1200 to 1500 nm long \cite{Angers}.

\begin{figure}[h!]
	\centering
	\includegraphics[width=\linewidth]{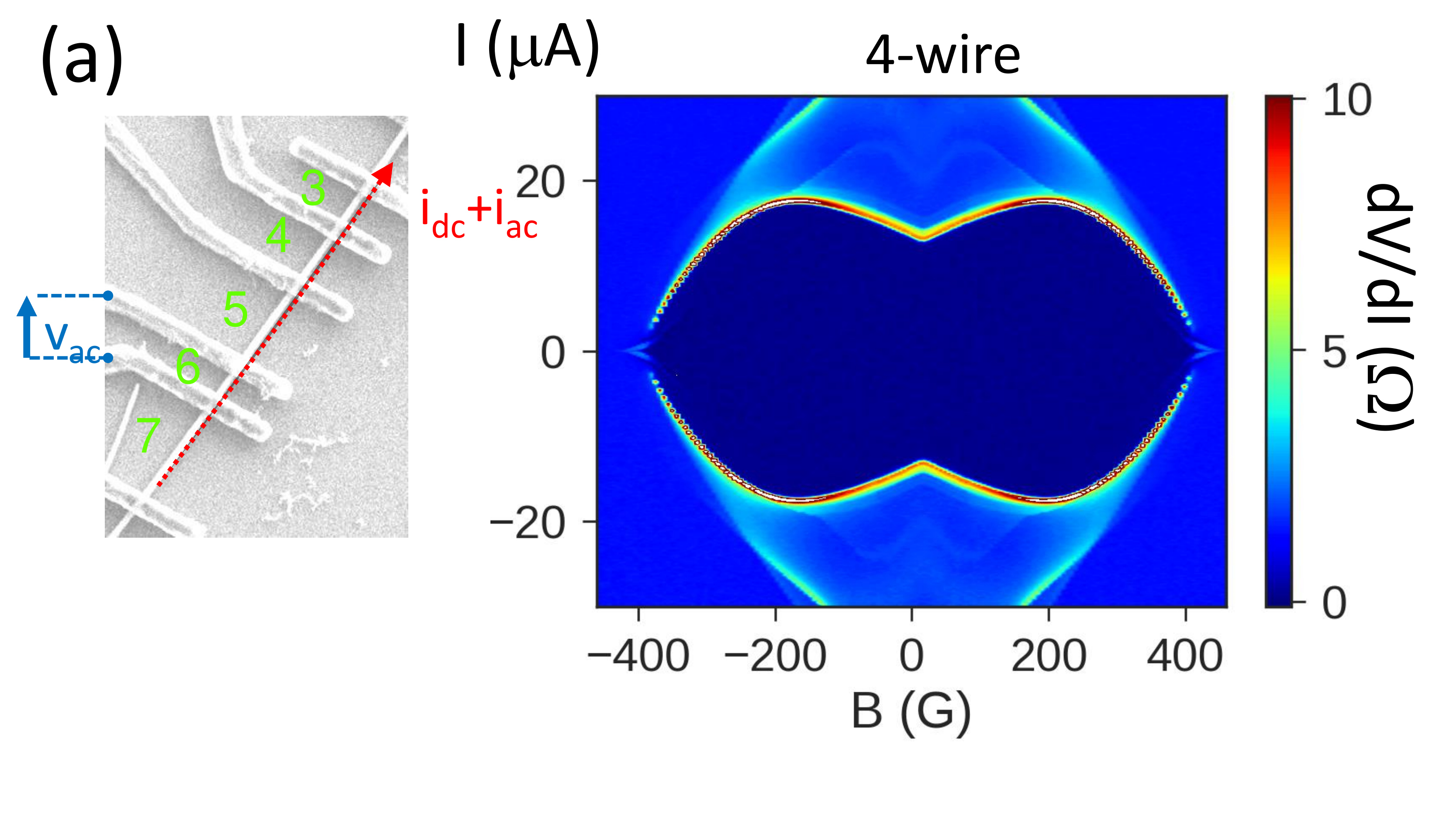}
	\includegraphics[width=\linewidth]{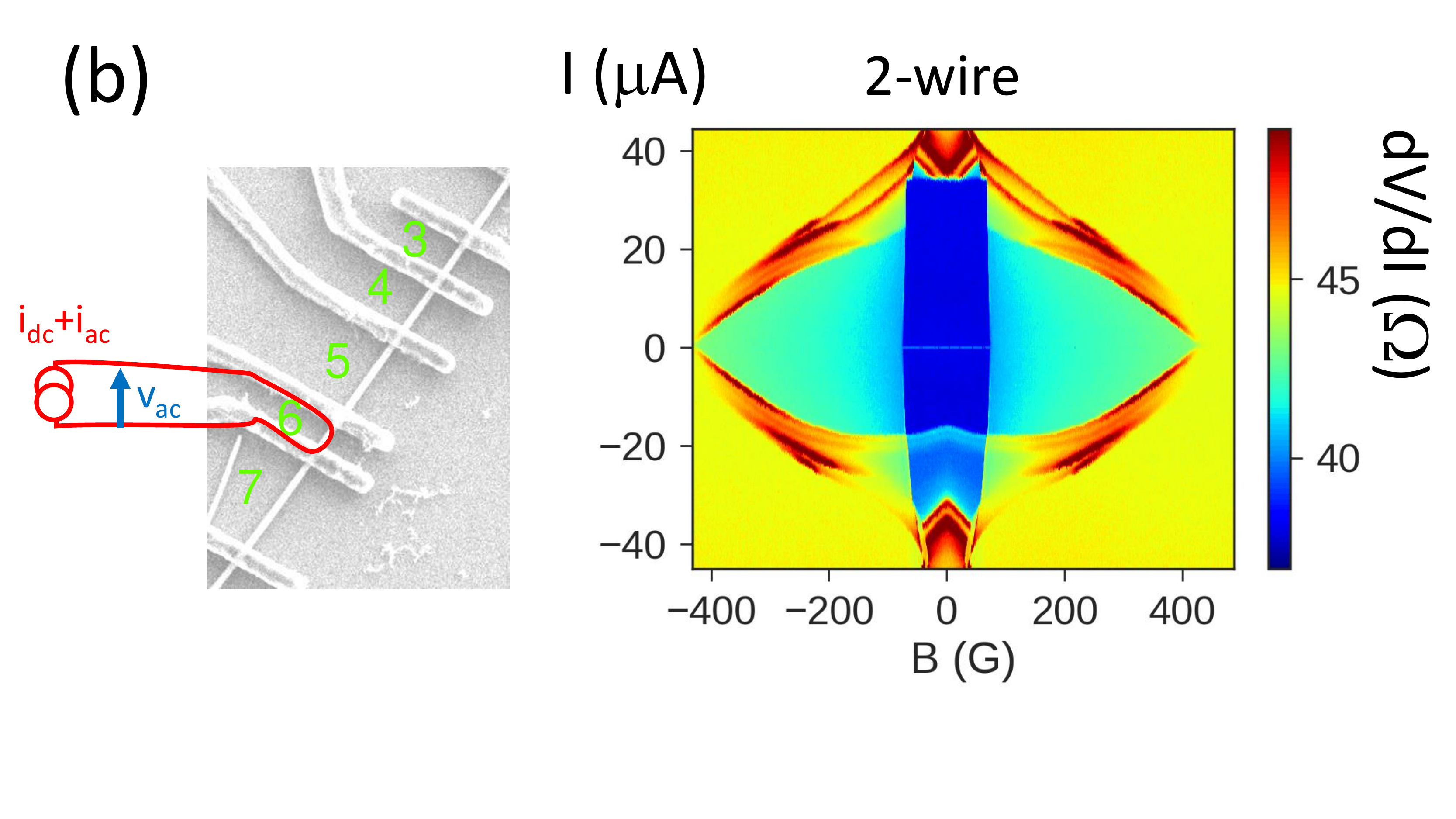}
	\caption{Colour-coded differential resistance $ \rm dV/dI=v_{ac}/i_{ac}$ of segment Ag10-6 with TiAl electrodes, in four-wire (A) and two-wire (B) measurement, as a function of dc current and magnetic field. In the four-wire configuration, the switching current is equal to the retrapping current, indicating sample heating, and the reentrant feature with magnetic field is clear. By contrast, in the two-wire configuration, the switching current at low field is up to twice the retrapping current, indicating that the sample is better thermalized. The two-wire resistance measures all leads up to the macroscopic bonding pads, so that the critical field $H_{c,b}$ of the bulk contact pads is visible at 75 G, which corresponds to one flux quantum through an area of size $\xi_{Al}^2$, and the proximity effect disappears at the critical field of the electrodes, 400 G, which corresponds to one flux quantum through an area of size $\xi_{Al}W$, where $W=170~\rm nm$ is the Al electrode width. 
	\label{Reentrance_2W4W}}
\end{figure}

\begin{figure}[h!]
	\centering
	\includegraphics[width=\linewidth]{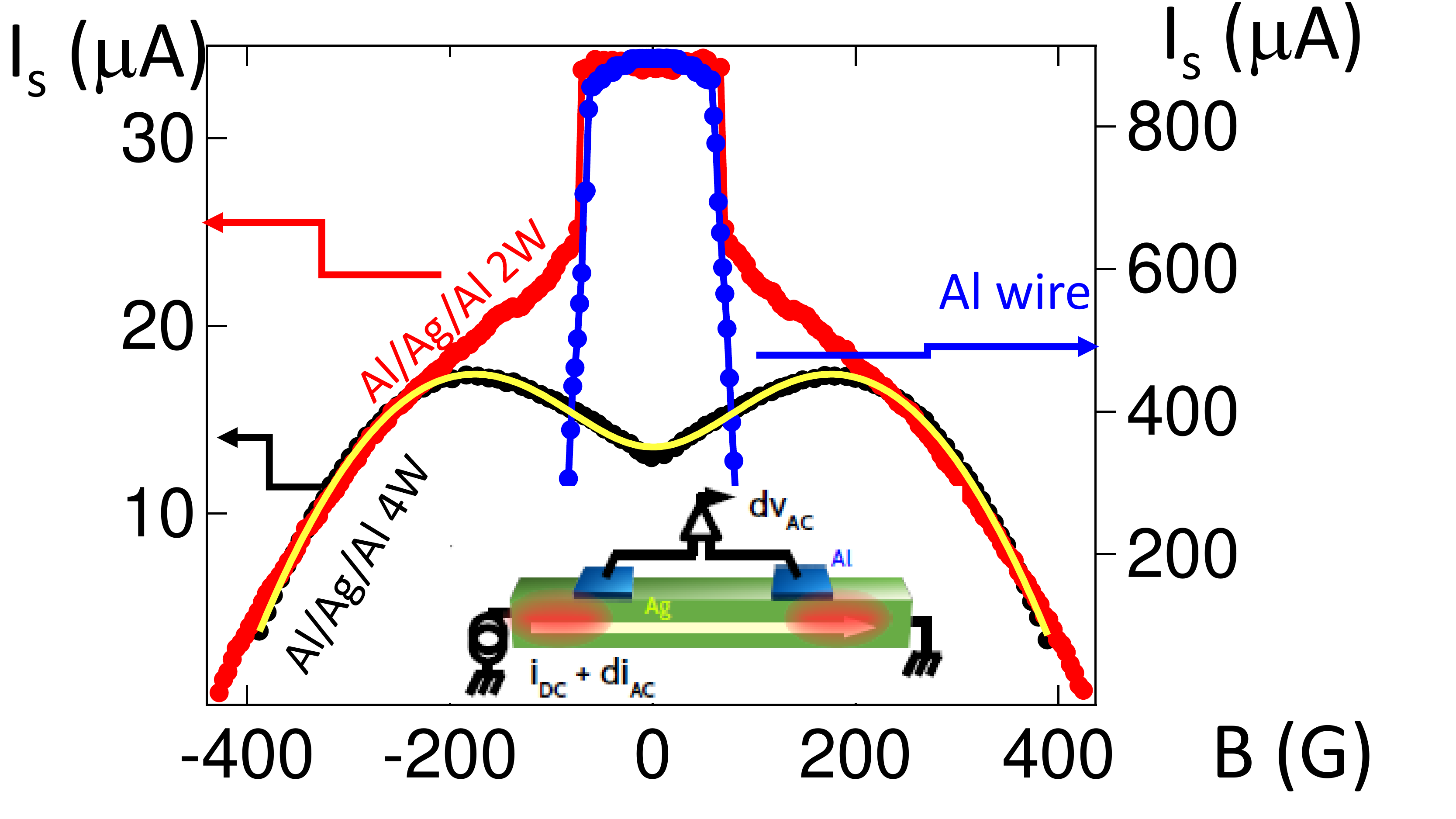}
	\caption{Comparison of the switching current of segment Ag10-6 with TiAl electrodes, in four-wire (black curve) and two-wire (red curve) measurement, with the switching current of an Al wire of same dimensions as the TiAl electrodes used to contact the Ag wires (blue curve, vertical scale on the right). No reentrant behaviour is observed in the Al wire field dependence, indicating that the reentrance is a feature of the proximity effect and not of the sole Al nanowire. The sharp cutoff starting around 75 G corresponds to the bulk Al's critical field $H_{c,b}$, given by one flux quantum through an area of $(500~\rm nm)^2$, in accordance with the superconducting coherence length of a disordered aluminum thin film (estimated in the text).
	 The yellow curve is the fit of the 4 wire measurement to expression (3) of the appendix describing the field-reentrance behavior.}		
		\label{comp_filAl}
		
\end{figure}
To investigate this intriguing behaviour of the Al/Ag/Al junctions, we compared two-wire and four-wire measurements of a given junction.  Figure \ref{Reentrance_2W4W} displays the differential resistance versus current and magnetic field of a given junction in two different configurations. The first is a four wire configuration, shown in Fig. \ref{Reentrance_2W4W}a, in which the current is injected via the Ag nanowire outside the segment, and the voltage drop is recorded between the superconducting probes connecting the segment. The reentrant feature is clear, both in the switching and retrapping current. It is also notable that in this four wire configuration, the switching current (the current above which the resistance goes from zero to the normal state value) is the same as the retrapping current (the current below which the resistance goes from the normal state value to zero). By contrast, in the two-wire configuration, shown in fig. \ref{Reentrance_2W4W}b, the same, superconducting, electrodes are used to inject the current and to measure the voltage drop. Strikingly, only the retrapping current displays a reentrance in that case, and the switching current is higher than the retrapping current. 

This behavior recalls that of field-enhanced superconductivity seen in extremely narrow (less than 10 nm) superconducting wires of amorphous MoGe and Nb \cite{Bezryadin,Feigelman}. Field enhanced superconductivity is attributed to the beneficial effect of the magnetic field in freezing out the spin of magnetic impurities, which in zero field act as inelastic scattering centers and decrease the critical current. 
We believe that the mechanism at play in our case of SNS junctions is different since magnetic impurities are Kondo screened in Al \cite{Mermin}. Moreover, we found no field-enhanced superconductivity (field reentrance) in test Al wires of the same thickness and width (170 nm) as the Al contacts used in the S/Ag/S junctions that displayed the anomalous field reentrance, as shown in Fig. \ref{comp_filAl}.

A qualitative description of the physics that could be at work is suggested by the difference between the two wire and four wire measurement. In the four wire configuration, the current is injected via the Ag nanowire. It is thus possible that there is a non-fully proximitized region in that wire outside the segment being measured. That would cause not only paired electrons, but also hot quasiparticles to be injected in the Ag section we measure. In zero field, those quasiparticles could not relax, in particular due to the hard superconducting gap of the Al electrodes, and would thus "poison" the junction, causing the switching current to be smaller. A magnetic field however softens the superconducting gap, i.e. induces some quasiparticle density of states in the gap,  so the injected quasiparticles could thermalize in the superconducting electrodes at lower temperature than at zero field, which would increase the critical current. 
A similar mechanism could explain the field dependence of the retrapping current, since the retrapping current depends on the cooling possibilities of hot quasiparticles \cite{Courtoisretrapping}.
 In the appendix, we propose a model of how the Joule power dissipated in a small normal portion of the Ag wire reduces the switching current, by equating the heat produced with the reduction of Josephson energy. The fit using this model is shown in Fig. \ref{comp_filAl}, and describes the data well. 

In contrast with the 4 wires switching current, the switching current in the 2 wire configuration, in which the current is injected in the Ag segment via superconducting Al electrodes, is maximum at zero magnetic field. It jumps down at 75 G, the critical field $H_{c,b}$ of the bulk contact pads, as visible in the (roughly 1 mA) critical current of an Al wire of similar dimensions, see Fig. \ref{comp_filAl}. This jump is followed at higher field by a smooth parabolic decay which coincides above 200 G with the field dependence of the 4 wire configuration. Both switching currents (2 and 4 wire configurations) go to zero above 400 G, the critical field of the  100 nm-wide aluminum electrodes that contact the Ag wires.
Since the Al bulk contact pads are more than several hundred microns away from the SNS junctions, it is surprising that their field transition should influence the switching current. Our understanding is that the switching current decreases above $H_{c,b}$ because hot quasiparticles are generated in the normal part of the bulk electrode and injected in the narrow part which is still superconducting.  

It is interesting that the two configurations have strikingly different magnetic field dependences, even though both behaviours are due to hot quasiparticles generated in the aluminum electrodes. The difference is due to the fact that the quasiparticles are generated either in the bulk electrodes (2 probe configurations) or close to the NS interface (4 probe configuration). The strong effect of these quasiparticles is due to the uncommonly small inelastic electron-phonon scattering rate in Al at low temperature \cite{Prober}, which implies that cooling is controled by the quasiparticles in aluminum. 

		\section{Conclusion}

	We have presented an extensive investigation of Ag nanowires-based Josephson junctions, with Aluminum as the superconducting material,  from the short to the long junction regime. Our experiments, and specifically the unusual field dependence of the critical current, strikingly demonstrate the importance of the thermalization of quasiparticles at the NS interface in systems where thermalization must proceed through aluminum electrodes.

\section{Acknowledgments}

We acknowledge Mathieu Kociak for the TEM imaging, Meydi Ferrier for insightful suggestions on the manuscript and funding from ANR of projects MASH (ANR-12-BS04-0016), MAGMA (ANR-16-CE29-0027), JETS (ANR-16-CE30-0029), and DIRACFORMAG (ANR-14-CE32-0003), and Labex PALM.

\section{Appendix}

\subsection{Parameters of TiAl/Ag/TiAl junction Ag6}	
\begin{table}[htbp]
	\begin{tabular}{|l|r|r|r|r|r|r|r|}
		\hline
		Segment  & \multicolumn{1}{l|}{Ag6-KM} & \multicolumn{1}{l|}{Ag6-DU} & \multicolumn{1}{l|}{Ag6-DZ} & \multicolumn{1}{l|}{Ag6-ZX}  \\ \hline
		Material & \multicolumn{1}{l|}{TiAl} & \multicolumn{1}{l|}{TiAl} & \multicolumn{1}{l|}{TiAl} & \multicolumn{1}{l|}{TiAl} \\ \hline
		$\Delta (\mu eV)$ & 200 & 200 & 200 & 340  \\ \hline
		$Lengths (\mu m)$ & 0.78 & 1.06 & 5.07 & 0.89  \\ \hline
		$R_{tot} (\Omega)$ & 4 & 4 & 10 & 4 \\ \hline
		$R_c (\Omega)$ & $1.8 \pm 1$ & $1.8 \pm 1$ & $1.8 \pm 1$ & $1.8 \pm 1$ \\ \hline
		$	R_N (\Omega) $& 1.30 & 1.30 & 7.30 & 1.30 \\ \hline
		$I_c (\mu A)$ & 8.5 & 7.5 & 0.39 & 6.1\\ \hline
		$e R_N I_c (\mu eV)$ & 34 & 30 & 3.9 & 24.4 \\ \hline
		$e R_N I_c / \Delta$& 0.17 & 0.15 & 0.02 & 0.07  \\ \hline
		$E_{Th}(\mu eV)$ & 32.36 & 17.52 & 0.77 & 24.85  \\ \hline
		$10.82 E_{Th}/\Delta $& 1.75 & 0.95 & 0.04 & 0.79  \\ \hline
		$k_B T / E_{Th}$ & 0.29 & 0.54 & 12.31 & 0.38   \\ \hline
		$e R_N I_c / E_{Th}$ & 1.05 & 1.71 & 5.09 & 0.98  \\ \hline
		$\xi$  (nm) & 313 & 313 & 313 & 313 \\ \hline
		$L/\xi$ & 2.49 & 3.39 & 16.20 & 2.84  \\ \hline
		$R_c/2 R_N$ & 1.04 & 1.04 & 0.18 & 1.04  \\ \hline
	\end{tabular}
	\caption{Parameters of the proximity effect induced in four segments of wire Ag6 connected to TiAl electrodes (see SEM image in main text Fig. 2b). }
	\label{Tableau}
	\label{}
\end{table}

\subsection{Model for field reentrance of switching current}

\begin{figure}[h!]
	\hspace*{-0.0cm}
	%\centering
	\includegraphics[width=9 cm]{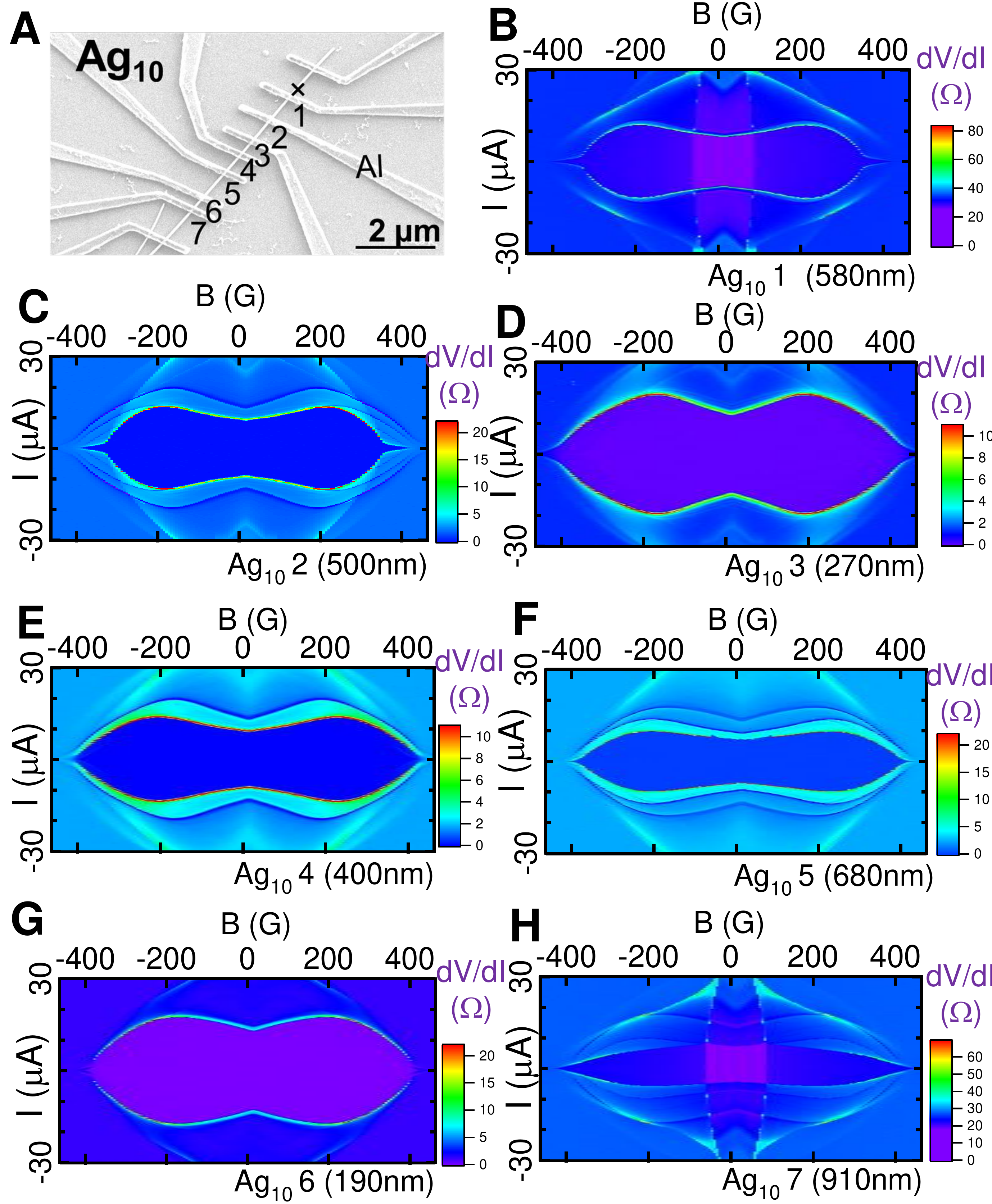}
	\caption{Colour-coded differential resistance of seven segments of Ag wires connected to TiAl electrodes, as a function of bias current and magnetic field. Segments 1 and 7 and measured in a three wire configuration, segments 2 to 6  are measured in a four-wire configuration. The reentrant field dependence is clear.}
	\label{SMReentranceAl}
\end{figure}

In the following, we propose a model of how the Joule power dissipated in a small normal portion of the Ag wire reduces the switching current, by equating the heat produced with the reduction of Josephson energy. 

We assume that a small normal part in the Ag wire near the Al contact causes Joule heating with power $R_{\rm qp}i(B)^2$. This power is converted over the quasiparticle thermalization time $\tau_{\rm qp}$, into an energy  that reduces the Josephson energy: 
\begin{equation}
R_{\rm qp} i^2(B) \tau_{\rm qp}=\phi_0\left(i_0(B)-i(B)\right),
\end{equation}
where $\tau_{\rm qp}$ is the characteristic electron-electron interaction time, and $\phi_0=\hbar/2e$ . 
The solution is 
\begin{equation}
%  i(B) = i_0(B)\left(\frac{-1+\sqrt{1+4 \kappa^2\tilde\tau_{qp}(B)}}{2 \kappa^2\tilde\tau_{qp}(B)}\right)
\end{equation}
\begin{equation}
i(B) =\frac{\phi_0}{2R_{\rm qp}}\frac{1}{\tau_{\rm qp}(B)}\left(-1+\sqrt{1+4 \frac{R_{\rm qp}}{\phi_0}i_0(B)\tau_{\rm qp}(B)}\right).
\end{equation}
%\begin{equation}
%i(B) = i_0(B)\left(\frac{-1+\sqrt{1+4 \kappa^2\tilde\tau_{R}(B)}}{2 \kappa^2\tilde\tau_{R}(B)%}\right)
%\end{equation}
%where I introduce the dimensionless magnetic field dependent recombination time $\tilde\tau_{\rm R}(B) = \tau_{\rm R}(B) / \tau_{\rm R}(0)$, and the dimensionless quantity $\kappa^2 = R_{\rm qp} \tau_{\rm R}(0)i_0/\phi_0$. 
%Ri^2 \tau_{\rm R} = \phi_0 \left( i_0 - i \right)
Here $i_0(B)$  is the critical current that would be measured with no Joule effect. Its decay with magnetic field is due to both the field-induced depairing in the superconducting  contacts, and the field-dependent decay of the supercurrent through an SNS junction \cite{Chiodi,Anthore}. A simple approximation is:
\begin{equation}
i_0(B) = i_0(0)\left( 1 - \left(\frac{B}{B_0}\right)^2\right),
\end{equation}
where $B_0$ is the characteristic magnetic field of the decay.
The quasiparticle thermalization rate, in contrast, increases with magnetic field, as the quasiparticle density of states in the superconductor increases. The simplest approximation is to take a rate proportional to the square of the magnetic field,

\begin{equation}
\begin{aligned}
\tau_{\rm qp}(B)^{-1} = \tau_{\rm qp}(0)^{-1}\left(1 + \left(\frac{B}{B_1}\right)^2\right).
\end{aligned}
\end{equation}

A fit to the data is shown in Fig. 6 of the main text, and qualitatively describes the data well. The fit parameters are $B_0=415~G$, the high field decay scale of the supercurrent;  $B_1=106~ G$, the field scale over which the quasiparticle thermalization time varies with field, leading to the reentrant part of the curve (positive $i(B)$ slope); $i_0=30.3~\mu A$ and $\frac{\phi_0}{2R_{\rm qp}\tau_{\rm qp}(0)}=10.95 ~\mu A$. Since the zero field thermalization time $\tau_{\rm qp}(0)$ is about $1-10\cdot 10^{-9}\,{\rm s}$ in Al \cite{Prober}, this yields $R_{\rm qp} ~0.1 \Omega$ as the value of the resistance causing the Joule power dissipation.

\begin{figure}[h!]
	\centering
	\includegraphics[width=\linewidth]{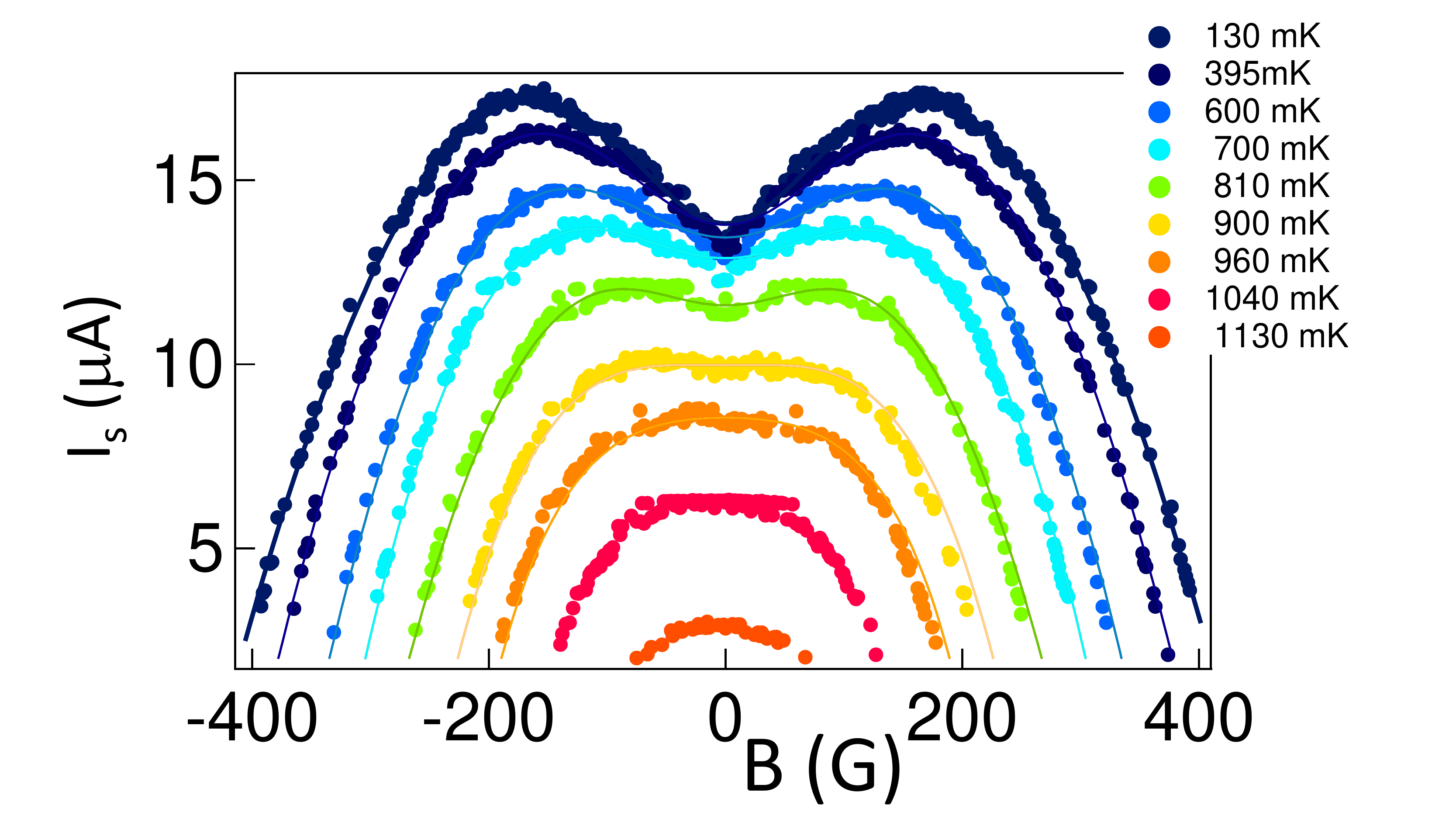}
	\caption{Variations with temperature of the reentrant feature in the switching current versus field curves of segment Ag10-6. In accordance with our understanding of the non-monotonous field dependence of the switching current as due to the heating effects of injected quasiparticles, the reentrance amplitude decreases as the sample temperature increases, since the cooling power of phonons increases with temperature. The experimental curves (markers) are taken at temperatures between 130 and 1130 mK, as indicated.  Fits of the curves to equation (3) are plotted as continuous lines. The curve shape is well described by the model, although the lowest temperature curves display a somewhat sharper zero field cusp than described by the model. }		
	\label{Ic_B_T}
	
\end{figure}

\end{document}